\documentclass[aps,pra,twocolumn]{revtex4}
\usepackage{graphics}
\usepackage{graphicx}

\begin{document}


\title{Destruction of Anderson localization in quantum nonlinear Schr\"odinger lattices}


\author{Alexander~V.~Milovanov${}^{1,2,4}$ and Alexander~Iomin${}^{3,4}$}

\affiliation{${}^1$ENEA National Laboratory, Centro~Ricerche~Frascati, I-00044 Frascati, Rome, Italy}
\affiliation{${}^2$Space Research Institute, Russian Academy of Sciences, 117997 Moscow, Russia}
\affiliation{${}^3$Department of Physics, Technion$-$Israel Institute of Technology, 32000 Haifa, Israel}
\affiliation{${}^4$Max-Planck-Institut f\"ur Physik komplexer Systeme, D-01187 Dresden, Germany}




\begin{abstract} The four-wave interaction in quantum nonlinear Schr\"odinger lattices with disorder is shown to destroy the Anderson localization of waves, giving rise to unlimited spreading of the nonlinear field to large distances. Moreover, the process is not thresholded in the quantum domain, contrary to its ``classical" counterpart, and leads to an accelerated spreading of the subdiffusive type, with the dispersion $\langle(\Delta n)^2\rangle \sim t^{1/2}$ for $t\rightarrow+\infty$. The results, presented here, shed new light on the origin of subdiffusion in systems with a broad distribution of relaxation times.
\end{abstract}


\maketitle

Anderson localization is the halt of diffusion of waves in a disordered medium. The phenomenon was predicted by Anderson in 1958 \cite{And} and has been extensively studied ever since, leading to outstanding advances in solid state physics, photonics and acoustics. The localization occurs because the wave process is scattered by inhomogeneities of the medium, causing the different components of the wave field to interfere with itself. At high disorder, the counter-propagating waves from reciprocal multiple scattering paths form a standing wave, thus hampering transport to large distances. This phase transition from diffusive transport to localization accommodates any type of classical or quantum wave, provided just that the mean free path becomes comparable to the wavelength. 

In a basic theory of disordered systems, the continued interest in Anderson localization was fueled by the suggestions, e.g., Refs. \cite{Sh93,PS}, that a weak nonlinearity can destroy the localized state and that the ensuing loss of localization is, in its turn, a critical phenomenon. That means that above a certain threshold strength of nonlinear interaction the nonlinear field may spread across the lattice to large distances (in spite of the underlying disorder) and is Anderson localized in the presence of these nonlinearities otherwise. Theoretically, the likely destruction of Anderson localization in nonlinear regimes has been studied in the fashion of the Gross-Pitaevskii equation with random potential \cite{Sh93,PS,Flach,Skokos,Fishman,Erez}. A modified perturbation theory addressing the nonlinear terms has been developed \cite{Fishman} and extensive numerical simulations have been carried out \cite{Flach,Skokos,Fishman}. A subdiffusive scaling for the onset spreading has been introduced and numerically measured \cite{PS,Flach}. Recently, a nonperturbative approach to the nonlinear Anderson dynamics has been proposed based on topological approximations, using random walks and the formalism of fractional diffusion equation \cite{Iomin,EPL,PRE14}. 

Even so, the theoretical investigations so far have assumed that the field is ``classical"$-$i.e., corresponds to the Gross-Pitaevskii mean field theory$-$making it possible to neglect the quantum tunneling effects and characterize the onset spreading in terms of the familiar picture of transition to chaos \cite{ZaslavskyUFN}. As will be shown in the present analysis, these assumptions are unnecessarily restrictive and tend to underestimate the transport. 

In this Report, we predict a subdiffusive law of spreading in a fully quantum case beyond the usual mean-field approximations. {\it This uses the idea of clustering of unstable modes in wave number space.} Our findings shed new light on the origin of anomalously slow diffusion observed in systems with a broad distribution of relaxation times \cite{Sokolov,Klafter}. We consider the problem of dynamical localization of waves in a quantum nonlinear Schr\"odinger equation (QNLSE) with random potential, i.e.,
\begin{equation}
i\hbar\frac{\partial\hat\psi_n}{\partial t} = \left(\hat{H}_L + \beta \hat\psi_n\hat\psi_n^{\dag}\right)\hat\psi_n,
\label{1} 
\end{equation}
where $\hat\psi_n = \hat\psi (n, t)$ is an operator wave function and is defined on a grid with the discrete coordinate $n$; the superscript $^{\dag}$ on $\hat\psi_n^{\dag}$ denotes Hermitian conjugate;
\begin{equation}
\hat{H}_L\hat\psi_n = \varepsilon_n\hat\psi_n + V (\hat\psi_{n+1} + \hat\psi_{n-1})
\label{2} 
\end{equation}
is the Hamiltonian of a linear problem in the tight binding approximation; $\beta$ ($\beta > 0$) characterizes the strength of nonlinearity; on-site energies $\varepsilon_n$ are randomly distributed with zero mean across a finite energy range; $V$ is hopping matrix element; and the total probability is normalized to $\sum_n \hat\psi_n\hat\psi_n^{\dag} = 1$. In what follows, $\hbar = 1$ for simplicity. For $\beta\rightarrow 0$, QNLSE with the Hamiltonian in Eq.~(\ref{2}) provides a quantum representation of the linear Anderson model as of Ref. \cite{And}. The eigenstates, $\phi_{n,k}$, of the linear problem are defined through $\hat H_L \phi_{n,k} = \omega_k \phi_{n,k}$ and constitute a full basis of mutually orthogonal functions with the eigenfrequencies $\omega_k$, $k=1,2,\dots$ Note that all eigenstates $\phi_{n,k}$ are exponentially localized in the linear phase, i.e., no spreading occurs for $\beta = 0$. 

To obtain the law of spreading in the nonlinear phase, it is convenient to use the basis of localized eigenstates $\phi_{n,k}$ of the linear problem with $\beta = 0$. Expanding the operator wave function $\hat\psi_n$ over the basis functions, we write $\hat\psi_n = \sum_m \hat a_m (t) \phi_{n,m}$, and similarly for $\hat\psi_n^{\dag}$, i.e., $\hat\psi_n^{\dag} = \sum_m \hat a_m^{\dag} (t) \phi^*_{n,m}$. Without loss of generality, we consider the eigenfunctions $\phi_{n,k}$ being normalized to unity, with the orthonormality condition $\sum _n \phi^*_{n,k_1}\phi_{n,k_2} = \delta_{k_1,k_2}$. Here, $\delta_{k_1,k_2}$ is Kronecker's delta and star denotes complex conjugation. $\hat a_m (t)$ and $\hat a_m^{\dag} (t)$ are, respectively, the annihilation and the creation bosonic operators obeying the natural commutation rule $[\hat a_{m_1} (t), \hat a_{m_2}^{\dag} (t)] = \delta_{m_1,m_2}$. With the aid of $\sum _n \phi^*_{n,k_1}\phi_{n,k_2} = \delta_{k_1,k_2}$ one sees that $[\hat \psi_{i}, \hat \psi_{j}^{\dag}] = \hat \psi_{i} \hat \psi_{j}^{\dag} - \hat \psi_{j}^{\dag}\hat \psi_{i} = \delta_{i,j}$ for all pairs of indices $i,j$. The total probability being equal to 1 implies $\sum_n \hat\psi_n\hat\psi_n^{\dag} = \sum_m \hat a_m (t) \hat a_m^{\dag} (t) = 1$. Multiplying both sides of Eq.~(\ref{1}) by $\phi^*_{n,k}$, then summing over $n$ and making use of the orthonormality condition, one obtains equations for the amplitudes $\hat a_{k} (t)$:   
\begin{equation}
i\dot{\hat a}_k - \omega_k \hat a_k = \beta \sum_{m_1, m_2, m_3} V_{k, m_1, m_2, m_3} \hat a_{m_1} \hat a^{\dag}_{m_2} \hat a_{m_3},
\label{4} 
\end{equation}
where $\omega_k$, $k=1,2,\dots$, are eigenvalues of the linear problem, the coefficients $V_{k, m_1, m_2, m_3}$ are defined by
\begin{equation}
V_{k, m_1, m_2, m_3} = \sum_{n} \phi^*_{n,k}\phi_{n,m_1}\phi^*_{n,m_2}\phi_{n,m_3},
\label{5} 
\end{equation}
and we have used dot to denote time derivative. Equations~(\ref{4}) correspond to a system of coupled nonlinear oscillators with the Hamiltonian 
\begin{equation}
\hat H = \hat H_{0} + \hat H_{\rm int}, \ \ \ \hat H_0 = \sum_k \omega_k \hat a^{\dag}_k \hat a_k,
\label{6} 
\end{equation}
\begin{equation}
\hat H_{\rm int} = \frac{\beta}{2} \sum_{k, m_1, m_2, m_3} V_{k, m_1, m_2, m_3} \hat a^{\dag}_k \hat a_{m_1} \hat a^{\dag}_{m_2} \hat a_{m_3}.
\label{6+} 
\end{equation}
Here, $\hat H_{0}$ is the Hamiltonian of non-interacting harmonic oscillators and $\hat H_{\rm int}$ is the interaction Hamiltonian. Note that we have included self-interactions into the definition of $\hat H_{\rm int}$. Each nonlinear oscillator with the Hamiltonian   
\begin{equation}
\hat h_{k} = \omega_k \hat a^{\dag}_k \hat a_k + \frac{\beta}{2} V_{k, k, k, k} \hat a^{\dag}_k \hat a_{k} \hat a^{\dag}_{k} \hat a_{k}
\label{6+h} 
\end{equation}
and the equation of motion 
\begin{equation}
i\dot{\hat a}_k - \omega_k \hat a_k - \beta V_{k, k, k, k} \hat a_{k} \hat a^{\dag}_{k} \hat a_{k} = 0
\label{eq} 
\end{equation}
represents one nonlinear eigenstate in the system $-$ identified by its wave number $k$, unperturbed frequency $\omega_k$, and nonlinear frequency shift $\Delta \omega_{k} = \beta V_{k, k, k, k} \hat a_{k} \hat a^{\dag}_{k}$. Non-diagonal elements $V_{k, m_1, m_2, m_3}$ characterize couplings between each four eigenstates with wave numbers $k$, $m_1$, $m_2$, and $m_3$. It is understood that the excitation of each eigenstate is none other than the spreading of the wave field in wave number space. If the field is spread across a large number of states $\Delta n\gg 1$, then the conservation of the probability $\sum_n \hat\psi_n\hat\psi_n^{\dag} \sim \int |\psi_n|^2 d\Delta n = 1$ implies that $|\psi_n|^2 \sim 1/\Delta n$. In the basis of linear localized modes, the evolution of the operators ${\hat a}_m (t)$ is controlled by the cubic nonlinearity $\dot{\hat a}_m \sim \beta \hat a_{m_1} \hat a^{\dag}_{m_2} \hat a_{m_3}$. The rate of excitation of the newly involved modes is $R \sim |\dot{\psi}_n|^2$ and proportional to the cubic power of the probability density, $|\psi_n|^2$. Taking the conservation of the probability into account, $R\sim 1/ (\Delta n)^3$. On the other hand, the number of the newly excited modes per unit time is $d/dt \times \Delta n$, making it possible to assess $d/dt \times \Delta n \sim 1/ (\Delta n)^3$. The latter condition is different from the corresponding condition used in Ref. \cite{PS} in that we do not assume that the spreading of the wave field is of the diffusive type; nor do we involve any sort of random-phase approximation justifying such an assumption. Indeed, in quantum dynamics, the notion of chaos loses its classical meaning \cite{Casati}. Therefore, the time derivative $d/dt$ is applied to $\Delta n$ itself$-$as dictated by Fermi's golden rule \cite{Golden}$-$and not to the square of $\Delta n$, as of Refs. \cite{PS,PRE14}, leading to a different law of spreading. Integrating over time, one sees that $(\Delta n)^4 \propto t$. Writing the coefficient in front of $t$ as $4A$, with $A\sim \beta^2$, one obtains a subdiffusive spreading     
\begin{equation}
(\Delta n)^2 = 2\sqrt{A}\times t^{1/2}.
\label{Sub} 
\end{equation}
The scaling in Eq.~(\ref{Sub}) corresponds to a faster process as compared to the law of spreading in the classically chaotic domain, $(\Delta n)^2 \sim t^{2/5}$ \cite{PS,PRE14}. The explanation lies in the fact that the quantum specific phenomena, such as tunneling between states, etc., naturally enhance the transport above the classically expected values. 

{\it Let us summarize:} Quantum transport may be much faster than the classic estimates would predict. In the case of QNLSE with a disordered potential, we find using the golden rule: $(\Delta n)^2 \propto t^{1/2}$. This scaling law agrees with the computer simulations of quantum diffusion in many body systems \cite{Flach_89}. It also agrees with the experimentally measured transport of nanoscale energy in molecular crystals and disordered thin films, Ref. \cite{Axel}.   

Let us now assess the dynamics of field spreading from the perspective of the second-order time derivative. For this, differentiate the equation $d/dt \times \Delta n = A / (\Delta n)^3$ with respect to time, then eliminate on the right-hand-side the first derivative $d/dt \times \Delta n$ with the aid of this equation. The end result is $d^2/dt^2 \times \Delta n = -3A^2 / (\Delta n)^7$. Rewriting the power-law dependence on the right-hand-side such that it takes the form of a ``gradient" in the $\Delta n$ direction, one gets
\begin{equation}
\frac{d^2}{dt^2}\times \Delta n = - \frac{d}{d \Delta n} \left[- \frac{A^2 / 2}{(\Delta n)^6}\right].
\label{Grad} 
\end{equation}
So, if $\Delta n$ is a position coordinate in wave number space, as in fact it is, then Eq.~(\ref{Grad}) is none other than the Newtonian equation of motion in the potential field 
\begin{equation}
W (\Delta n) = - \frac{A^2 / 2}{(\Delta n)^6}.
\label{Poten} 
\end{equation}
The potential function in Eq.~(\ref{Poten}) is immediately recognized as the attractive part of the celebrated Lennard-Jones potential \cite{Lennard}, which finds outstanding applications in molecular dynamics and quantum chemistry. As a result of this attraction, the newly excited modes will tend to form clusters$-$``molecules"$-$in wave number space; where they will be effectively trapped due to their nonlinear coupling. The comprehension of the attractive ``forces" between the components of the wave field will help to explain the deviation from the normal diffusion in the nonlinear Schr\"odinger dynamics. Indeed the transport is subdiffusive, i.e., $(\Delta n)^2 \sim t^{1/2}$, and not $\sim t$ as in the normal transport case, owing to the binding effect of the potential field of the Lennard-Jones type. We shall illustrate this property shortly.   

Multiplying both sides of Eq.~(\ref{Grad}) by the ``velocity", $d/dt \times \Delta n$, and integrating the ensuing differential equation with respect to time, after simple algebra one obtains
\begin{equation}
\frac{1}{2}\left[\frac{d}{dt}\times \Delta n\right]^2 - \frac{A^2 / 2}{(\Delta n)^6} = \Delta E,
\label{Ener} 
\end{equation}
where the first term on the left-hand-side has the sense of the kinetic energy of a ``particle" of unit mass moving along the $\Delta n$ coordinate, and the second term is its potential energy. It is shown using the equation $d/dt \times \Delta n = A / (\Delta n)^3$ that the kinetic energy in Eq.~(\ref{Ener}) compensates the potential energy {\it exactly}, that is, the full energy in Eq.~(\ref{Ener}) is zero, $\Delta E = 0$. More so, both the negative potential energy $W (\Delta n) \sim - 1 / (\Delta n)^6$ and the positive kinetic energy $\frac{1}{2}(d/dt \times \Delta n)^2 \sim 1 / (\Delta n)^6$ vanish while spreading. Both will decay as the inverse sixth power of the number of states and the ratio between them will {\it not} depend on the width of the field distribution. 

The full energy being equal to zero implies that the ``particle" in Eq.~(\ref{Ener}) is sitting on the separatrix $\Delta E = 0$; which naturally allows an escape path to infinity, hence unlimited spreading of the wave field regardless of the strength of nonlinearity. More so, as the particle propagates outward, its motion becomes intrinsically unstable (sensitive to fluctuations). This is because both the potential and the kinetic energies vanish for $\Delta n \rightarrow +\infty$, so very tiny perturbations due to for instance random noise, zero point fluctuations, quantum tunneling, etc. may drastically change the type of phase space trajectory. The result generally holds for dynamics near separatrices in large systems \cite{ZaslavskyUFN,ChV}. To this end, the fact that a given mode does or does not belong to a cluster of modes becomes essentially a matter of probability. 

To assess the probabilistic aspects of field spreading, let us assume that the fluctuation background is characterized by the effective ``temperature", $T$. So, the value of $T$ weighs all occasional perturbations to dynamics that might be influential near the separatrix. Then the probability for a given mode to quit the cluster after it has traveled $\Delta n$ sites on it is given by the Boltzmann factor $p_{\rm} (\Delta n) = \exp [W (\Delta n) / T]$. Here we measure temperatures in energy units, so we can set the Boltzmann constant to 1. Substituting $W (\Delta n)$ from the Lennard-Jones potential in Eq.~(\ref{Poten}), one finds 
\begin{equation}
p_{\rm} (\Delta n) = \exp [- A^2 / 2 T (\Delta n)^6] \approx 1 - A^2 / 2 T (\Delta n)^6,
\label{Escape} 
\end{equation}
where the exponential function has been expanded for $\Delta n \gg 1$. The probability to remain (``survive") on the cluster after $\Delta n$ space steps is $p^{\prime}_{\rm} = 1-p_{\rm} \approx A^2 / 2 T (\Delta n)^6$. Eliminating $\Delta n$ with the aid of Eq.~(\ref{Sub}), one obtains the probability to survive on the cluster for $t$ time steps, i.e.,  
\begin{equation}
p^{\prime}_{\rm} (t) \approx (\sqrt{A} / 16 T) \times t^{-3/2}.
\label{Survive} 
\end{equation}
Naturally, the survival probability in Eq.~(\ref{Survive}) can be interpreted as a waiting-time distribution $\chi (t) \propto (\tau / t)^{3/2}$, where $\tau \sim A^{1/3} / (16 T)^{2/3}$ is a normalization parameter. Note that the integral $\int t \chi (t) dt\sim t^{1/2}$ diverges for $t\rightarrow+\infty$, implying that the mean waiting time is infinite. 

In a basic statistical physics of random processes, the inclusion of the diverging mean waiting time leads to continuous-time generalizations of the Brownian random walk \cite{CTRW} and non-Markovian, non-Gaussian ventures into the familiar diffusion equation in the limit $t\rightarrow+\infty$. These naturally involve the exact form of the $\chi (t)$ dependence. For $\chi (t) \propto (\tau / t)^{1+\alpha}$, with $0 < \alpha < 1$, the asymptotic ($t\rightarrow+\infty$) transport equation deriving from these generalizations reads \cite{Sokolov,Klafter}  
\begin{equation}
\frac{\partial}{\partial t} f (t, \Delta n) = \frac{\partial^2}{\partial (\Delta n)^2} K_\alpha \frac{\partial}{\partial t}\int _{0}^{t} dt^{\prime} \frac{f (t^{\prime}, \Delta n)}{(t - t^{\prime})^{1-\alpha}},\label{FDE} 
\end{equation}
where $K_\alpha$ absorbs in one number the parameters of the transport process, $f (t, \Delta n)$ is the probability density to find the random walker at time $t$ at the distance $\Delta n$ from the origin, and we have chosen $t=0$ as the starting point for dynamics. Equation~(\ref{FDE}) is the much discussed {fractional} diffusion equation describing subdiffusion \cite{Sokolov,Klafter}. The second moment of the probability density grows as $\langle(\Delta n)^2\rangle \propto t^\alpha$ for $t\rightarrow+\infty$. One sees that the spreading problem for QNLSE joins into this picture of time fractional diffusion equation with $\alpha = 1/2$. 

It is worth emphasizing that the non-Markovian properties arise naturally through dynamics via the action of the Lennard-Jones potential causing attraction between the unstable modes. It is due to this attraction that the transport is slowed down below its diffusionlike values. Behind the subdiffusive character of the spreading is the nonlinear interaction between the modes; in particular, the four-wave interaction in Eqs.~(\ref{4}) generates a waiting-time distribution with the divergent mean, enabling non-Markovian dependencies in Eq.~(\ref{FDE}). 

Previous calculations of $\alpha$ used a completely different approach \cite{EPL,PRE14}. It was argued that the wave field was a mixture of oscillators in regular state and dephased state. It was then assumed that the transport may only occur between the oscillators in dephased state via a next-neighbor rule. Then the field could propagate to infinity only if the concentration of the dephased oscillators exceeds a certain threshold. This threshold concentration permitting transport to large distances is obtained as the percolation threshold on a Cayley tree. Focusing on the dynamical equations~(\ref{4}), when summing on the right-hand side, the only combinations of terms to be taken into account are, for the reasons of symmetry, $\hat a_{k} \hat a^{\dag}_{k} \hat a_{k}$ and $\hat a_{k-1} \hat a^{\dag}_{k} \hat a_{k+1}$. This yields
\begin{equation}
i\dot{\hat a}_k - \omega_k \hat a_k = \beta V_{k} \hat a_{k} \hat a^{\dag}_{k} \hat a_{k} + 2\beta V_k^\pm \hat a_{k-1} \hat a^{\dag}_{k} \hat a_{k+1},
\label{9} 
\end{equation} 
where we have also denoted for simplicity $V_k = V_{k, k, k, k}$ and $V_k^\pm = V_{k, k-1, k, k+1}$. Equations~(\ref{9}) define an infinite ($k= 1,2,\dots$) chain of coupled nonlinear oscillators where all couplings are local (nearest-neighbor-like). The mapping on a Cayley tree is obtained as follows. A node with the coordinate $k$ represents a nonlinear eigenstate, or nonlinear oscillator with the equation of motion~(\ref{eq}). There are exactly $z=3$ bonds at each node: one that we consider ingoing represents the creation operator $\hat a^{\dag}_{k}$, and the other two, the outgoing bonds, represent the annihilation operators $\hat a_{k-1}$ and $\hat a_{k+1}$ respectively. The percolation transition occurs at the critical concentration $p_c =1/(z-1) = 1/2$. For $p\rightarrow p_c$, the distribution of the dephased oscillators is self-similar (fractal). A random walker placed on the infinite cluster at percolation will exhibit anomalous dispersion $\langle(\Delta n)^2\rangle \sim t^{2/(2+\theta)}$ \cite{Gefen}, where $\theta$ is the index of anomalous diffusion and incorporates the topological characteristics of the cluster (such as connectivity, etc). The mean-field result, holding for percolation problem on a Cayley tree, is $\theta = 4$. Hence one predicts the dispersion $\langle(\Delta n)^2\rangle \sim t^{1/3}$, with $\alpha = 1/3$.

The discrepancy between the two models is not really surprising. In quantum diffusion, the assumption that the transport occurs between the next neighbors only is invalidated as soon as quantum tunneling is involved \cite{CSF}. Also the distinction between the states (regular {versus} dephased) is only possible in the ``classical" limit as it uses the classical criteria for the transition to chaos \cite{ZaslavskyUFN,ChV}. As a result, the classical picture of field spreading proves to be fairly different from a fully quantum description; in particular ({i}) it is thresholded, whereas the quantum scenario is not; ({ii}) leads to a slower transport, due to the next-neighbor limitation; and ({iii}) uses fractality and other self-similarity arguments, which find their theoretical justification in the thresholded character of spreading. 

The quantum picture of field spreading is based on Fermi's golden rule \cite{Golden} for quantum transitions between states. It is shown using QNLSE with disorder that the nonlinear coupling between the eigenstates generates some form of attractive potential in wave number space. This potential is of the Lennard-Jones type. The ensuing dynamics are such as to favor multiple trapping phenomena with a distribution of waiting times and the divergent mean waiting time. In a statistical perspective, this leads to a transport model involving non-Markovianity, based on an algebraically decaying memory response function. Mathematically, it corresponds to a description using fractional-derivative equations in the time domain. This connection to non-Markovian transport with long-time rests discussed here elucidates the theoretical significance of {\it fractional kinetics} \cite{Sokolov,Klafter} starting from the quantum grounds. A crossover between the quantum and the classical descriptions is predicted in terms of the $\alpha$ value. This crossover should show up through an increasing complexity of the transport process, leading to a thresholded (``critical") behavior \cite{PRE14,Asch} in the classical region of parameters. In view of the asymptotic ($t\rightarrow+\infty$) character of the transport, our conclusions support the finding of Ref. \cite{Berman} that a convergence of quantum solutions to the corresponding classical solutions may exist only for limited times. In general, we expect the effect of the attraction to favor clustering of the unstable modes. Clearly, quantum tunneling within the same cluster does not contribute to transport on the large scales. When the number of modes which belong to the same cluster becomes statistically significant, the clusters acquire signatures that enable to consider them as macroscopic states (``regular" {or} ``dephased"). Then quantum transitions between the different clusters may be described in some approximation as a next-neighbor random walk on a fractal lattice at percolation. This recovers the critical transport regimes already discussed in Refs. \cite{EPL,PRE14}. 

This work was supported in part by the Max-Planck-Institute for the Physics of Complex Systems (Dresden, Germany), by the Israel Science Foundation, and by the Eurofusion grant AWP17-ENR-ENEA-10.

%
%
%
%


\end{document}